\documentclass[12pt,preprint]{aastex}

\begin{document}
\title{An Ultra-luminous X-Ray Object with a 2 hour period in M51}

\author{Ji-Feng Liu, Joel N. Bregman, Jimmy Irwin, and Patrick Seitzer}
\affil{Astronomy Department, University of Michigan, MI 48109}

\begin{abstract}
Ultra-luminous X-Ray Objects are off-nucleus point sources with
$L_X=10^{39}$-$10^{41}$ erg/s but the nature of such systems are largely
unidentified.  Here we report a 2.1 hour period observed in a Chandra ACIS
observation for ULX M51 X-7, which is located on the edge of a young star
cluster in the star forming region in a spiral arm. In two ACIS observations
separated by one year, the ULX changed from a high-hard to a low-soft spectral
state, in contrast to most Galactic low mass X-ray binaries. Based on its
period and spectral behaviors, we suggest that this ULX is a low mass X-ray
binary system, with a dwarf companion of 0.2-0.3 $M_\odot$ and a compact
accretor, either a neutron star or a black hole, whose mass is not well
constrained. Relativistic beaming effects are likely involved to produce the
observed high X-ray luminosities given its low accretion rate as inferred from
a sustainable accretion scenario via Roche lobe overflow.

\end{abstract}


\section{Introduction}

X-ray emission from active galactic nuclei and quasars, powered by accretion
onto nuclear super-massive black holes, are usually $>10^{41}$ erg/s, while for
the stellar binary systems in the Milky Way, the luminosities are typically
$10^{33}$-$10^{38}$ erg/s.  X-ray objects with luminosities of
$10^{39}$-$10^{41}$ erg/s have been observed in some external galaxies as
off-nucleus point sources by X-ray satellites such as EINSTEIN, ROSAT, ASCA,
and recently Chandra X-ray Observatory (e.g., Fabbiano 1989; Marston et al.
1995; Zezas et al.  2002).  These objects have several names, a popular
designation being ``Ultra-luminous X-ray Objects'', or ULXs.

One suggestion for the nature of these ULXs is that they are binary systems
with $10^3$-$10^4$ $M_{\odot}$ black holes as primaries (Colbert and Mushotzky
1999). Such black holes, if they exist, are the missing links between stellar
mass black holes and super-massive black holes in the nuclei of galaxies. This
suggestion is consistent with the X-ray spectral  of some ULXs (Makishima et
al.  2000).  However, the formation of such massive black holes is not
predicted by stellar evolution theory and it may be impossible to form such
objects even in dense star clusters.  Alternatively, these sources may be
stellar mass black holes or neutron stars whose emission is beamed, thereby
representing micro-quasars (King et al. 2001; King 2002). If the emission is
beamed, the intrinsic luminosities become sub-Eddington, and there are known
examples of beamed Galactic X-ray sources (c.f., Mirabel \& Rodriguez 1999). It
is also possible that the luminosities are truly super-Eddington, for example,
obtainable from accretion disks with radiation-driven inhomogeneities
(Begelman, 2002).

To identify which of the above suggestions is correct, one natural and fruitful
way is to associate them with known classes of objects by comparing spectral
behaviors, short-term and long-term variabilities, etc. Spectral analyses of
ULXs, not without problems, generally support the mass-accreting black hole
scenarios (e.g., Makishima et al. 2000).  Spectral transitions were observed
for two ULXs in IC 342 between high-soft and low-hard states, reminiscent of
some black hole X-ray binaries in our Galaxy (Kubota et al.  2001).  While many
ULXs are variable during individual observations, few periods have been
reported. So far, only a possible periodicity at several tens of hours of a ULX
in IC 342 has been reported (Sugiho et al. 2001). In this paper, we report a
2.1 hour period for a ULX in M51 (NGC 5194), M51 X-7 as designated in Roberts
\& Warwick 2000. This ULX is located on the outskirts of a young open cluster
on a spiral arm of M51.  In $\S$2, we present the two Chandra ACIS observations
utilized and the results. A period of 7620 seconds for the ULX is found from
the X-ray light, and a spectral transition is found between two Chandra
observations. In $\S$3, we discuss the implications of its period and spectral
evolution, and suggest that this ULX is a low mass X-ray binary system, with a
dwarf companion of 0.2-0.3 $M_\odot$ and a compact accretor.  Possible
mechanisms to produce the high X-ray luminosity is also discussed.  For the
distance to M51, we use 7.7 Mpc (Roberts \& Warwick 2000). 

\section{Data analysis and results}

Two Chandra ACIS observations of M51, ObsID 354 (15 kilo-seconds,
2000-06-20) and ObsID 1622 (27 kilo-seconds, 2001-06-23), were retrieved from
the archive.  The data were analyzed with  CIAO 2.2.1,  CALDB 2.12, XSPEC
11.2.0 and XRONOS 5.19. Pixel randomization of event positions was removed to
recover the original positions before CIAO task {\tt wavdetect} was used to detect
discrete sources on ACIS-S chips.  The accuracy of absolute positions for these
sources is expected to be $0\farcs6$ (Aldcroft et al. 2000).

The ULX M51 X-7 was observed in both data sets, with ACIS positions
differing by ($0\farcs5$, $0\farcs1$). For the true
position, we take the average, i.e., RA=13:30:0.99, Dec=47:13:43.9, with a
positional uncertainty of $0\farcs4$. Registering this position onto
an HST WFPC2 image, we find that it falls onto the edge of a star cluster in a
dusty star forming knot on a spiral arm of M51. Within its positional error
circle are a few stars fainter than 23 mag. The details of its potential
optical counterpart will be reported in a future paper. 

For timing and spectral analyses, we selected circular apertures with such
sizes that the number of source photons outside of the aperture is
approximately the number of background photons inside the aperture. The size
chosen is $4\farcs0$ for Observation 354, and $2\farcs0$ for Observation 1622.
The light curve of X-7 in the first ACIS observation (ObsID 354), defined by
about 550 photons, appears to have the structure of a single period, as shown
in Figure 1 in contrast with the background light curve.  When fitting a
periodic signal, we derive a period of 7620 seconds (using XRONOS/efsearch with
10 second resolution). The uncertainty in the period is approximately 500
seconds (a more accurate value is not possible due to the limited number of
periods involved).  To determine whether this could be an instrumental effect,
we carried out a number of tests.  We formed the light curves of all other
bright sources in the same ACIS field, but found no evidence for periods.  The
period cannot be caused by background variation, since the count rate of the
background varies by less than 10\% (and is not periodic), while the count rate
of X-7 varies by an amplitude of more than 50\%, for both soft photons ($<1.5$
keV) and hard photons ($>1.5$ keV). 

Dithering may cause spurious periods when a source is dithering across a CCD
node boundary or gap, falls off of the detector, or crosses bad pixels
periodically. In this observation, the dither period is 1000 seconds in the
Chip Y direction, and 700 seconds in the Chip X direction. The dither pattern is
a Lissajous figure, spanning 16 arc-seconds from peak to peak. The ULX falls on
node 2 of chip S3 at (556,459). It is 22 arc-seconds away from the node 1-2
boundary, and no bad pixels fall within 30 arc-seconds of the source. Thus it is
unlikely that the period is due to the dithering period.

In the second ACIS observation (ObsID 1622) that was taken 1 year after the
first, only 53 photons were collected for the ULX, too few to make a reliable
periodicity analysis. Note that the second observation was longer than the
first by a factor of about 1.8, so the photon flux in the 0.3-10keV decreased
by a factor of about 18 between these two observations. 

Photons in the 0.3-8 keV band are used to perform spectral fits for both
spectra, as shown in Figure 2. The CIAO thread {\tt acisspec} was used to
extract the spectra, and {\tt corrarf} was used to correct the continuous
degradation in the ACIS quantum efficiency. To get good $\chi^2$ statistics, we
grouped the first spectrum (ObsID 354) requiring a minimum of 20 photons per
bin. For the second spectrum (ObsID 1622), we required a minimum of 5 photons
per bin to get enough bins.  We tried absorbed (wabs) power-law (powerlaw),
absorbed multi-color disk (diskbb) and absorbed blackbody (bb) models on each
spectrum.  In Table 1 we list the best fit parameters, the errors for the
best-fit parameters are for the 90\% confidence level. For the first spectrum,
an absorbed blackbody model is not acceptable, but an absorbed power-law model
is acceptable, with a power-law photon index $\Gamma=1.25^{+0.24}_{-0.22}$; and
an absorbed multi-color disk model is also acceptable, with an inner edge
temperature $kT_{in}=2.10^{+0.63}_{-0.40}$ keV.  To test whether there is an
additional component due to a neutron star solid surface, we also fit an
absorbed (multi-color disk + 2 keV blackbody) model to the first spectrum. With
one more free parameter, this model improves the spectral fit marginally as
compared to an absorbed multi-color disk model, and results in a smaller inner
edge temperature, $kT_{in}=0.82^{+1.57}_{-0.30}$ keV. However, this additional
component does not make a better model than a simple absorbed power-law model.

For the second spectrum, an absorbed power-law model gives a power-law photon
index of $\Gamma>2.6$, an absorbed multi-color disk model gives an inner edge
temperature of $kT_{in}=0.29\pm0.09$ keV, while an absorbed blackbody
model is not acceptable.  As compared with the first spectrum of higher photon
flux, the second spectrum is much softer, and the best-fit absorbing column
density is higher too. For simple absorbed power-law models, we switched the
photon indices of the two spectra, and fitted a frozen  $\Gamma=2.6$ to the
first spectrum, and a frozen $\Gamma=1.25$ to the second spectrum, and found
that both models can be accepted only on the levels $<10^{-2}$.  Thus we
conclude that the first spectrum is truly harder than the second spectrum.
Absorbed power-law models with smaller frozen $n_H$ values were also fitted to
the second spectrum and gave worse fits, indicating that the high column
density in the second spectrum is real.

\section{Discussion}

Periods are not uncommon in Galactic X-ray point sources, such as cataclysmic
variables (CVs), low mass X-ray binaries (LMXBs) and high mass X-ray binaries
(HMXBs). HMXBs have periods of more than a few days, because they are usually
widely separated systems. A two hour period is common for CVs, where the
accretor and the donor are white dwarfs.  However, CV systems have typical
temperatures under $10^6$ K, with most of the radiation emitted in the
ultraviolet region and with X-ray luminosities below $10^{34}$ erg/s. LMXBs
usually have periods of 10+ hours, but they can have periods as short as one
hour, for example, 1624-490 has a period of 0.69 hours.  The source studied
here, with a period of two hours and a temperature of $\sim1$ keV, is most
similar to the low mass X-ray binaries. 

The LMXB picture of ULXs does not contradict its location in a star forming
region.  While observations show that ULXs preferentially occur in active star
forming regions (e.g.,Zezas et al. 2002), indicating their short lifetimes,
they nevertheless could be a short stage in the long lifetime of LMXBs. Indeed,
some ULXs have also been found outside of star forming regions, for example, in
gas poor elliptical galaxies such as NGC 1399 (Angelini et al. 2001), and in a
globular cluster on the outskirt of the bulge in edge-on spiral galaxy NGC 4565
(Wu et al.  2002).

If X-7 is a LMXB with an orbital period of about 2 hours, there are a number of
implications.  The secondary star would be filling its Roche Lobe and
undergoing mass loss to the primary, and from this picture, we can estimate the
mass of the secondary, $M_1$. For a mass ratio of $q=M_1/M_2<0.8$, the Roche
Lobe radius is $R_{cr}=0.46a({M_1 \over M_1+M_2})^{1/3}$ (Paczynski 1967), in
which $a$ is the separation between the donor and the accretor, and $M_2$ is
the mass of the accretor. Combined with Kepler's 3rd law, this leads to
$P_{orb}=8.9(R_1/R_\odot)^{3/2}(M_\odot/M_1)^{1/2}$ hours.  For a late-type low
mass star, if we adopt an analytical mass-radius relation
$R_1/R_\odot=M_1/M_\odot$ (Verbunt 1993), we obtain the mass of the donor to be
$M_1\approx0.23M_\odot$.  We also tried to obtain the mass of the donor by
adopting computed mass-radius relation from the Geneva stellar models (Lejeune
et al. 2001). We extended the star mass down to 0.1 $M_\odot$ by extrapolating
computed lgR-lgM relation for low mass stars ($0.8<M/M_\odot<0.9$). These
computed stellar models give an equivalent analytical relation of
$R_1/R_\odot\approx(M_1/M_\odot)^{1.2}$, and the resulted donor mass is about
0.3 $M_\odot$. Such a low mass secondary in M51 is impossible to identify
even with the Hubble Space Telescope. 

The mass of the accretor cannot be constrained by the period under our
assumptions. The accretor is most likely a neutron star, or a black hole , as a
compact object is needed to produce the observed hard spectra.  One can refine
this suggestion further by comparing the observed spectrum with models of
compact objects, as we try here.  Neutron star LMXBs and black hole LMXBs at
high luminosities might be distinguished by their spectral features (cf. Tanaka
\& Lewin 1995), since high luminosity neutron star LMXBs exhibit a multi-color
disk component and a blackbody component with a color temperature of $\sim2$
keV, which is interpreted as emission from the neutron star surface, while
black hole LMXBs do not show such a single-color blackbody component at any
luminosity level. As the luminosity decreases further, the spectrum becomes a
single power-law, and distinction between black hole and neutron star LMXBs
diminishes.  For X-7, the model with an additional 2 keV thermal component
improves the spectral fit marginally as compared to an absorbed multi-color
disk model, which may be very weak evidence that the accretor is a neutron
star. However, such a marginal improvement may be due to the additional freedom
of parameters, and does not really reflect the preference of a thermal
component.  Also, its trend of spectral transition from a hard high state to a
soft low one has some Galactic parallels, such as Nova Muscae 1991 (Ebisawa et
al. 1994), whose system is comprised of a low mass companion and a black hole
of $\sim6$ $M_\odot$. Such spectral transitions, as noted by Terishima \&
Wilson (2002), are in contradiction to most Galactic LMXBs which switch between
soft high and hard low states. Thus it is premature to identify it as a neutron
star or a black hole with current observations. 

One concern about this picture is that mass transfer from the less massive
donor to the accretor will widen the orbit and enlarge the Roche lobe with
respect to the donor, and thus prevent further mass transfer via Roche lobe
overflow. This can be avoided if the orbit angular momentum can be dissipated
effectively by gravitational radiation, with the rate $-(\dot{J}/J)_{GR}={32G^3
\over 5c^5}M_1M_2(M_1+M_2)/a^4$. The separation $a$ evolves according to the
equation $\dot{a}/a = 2\dot{J}/J-2(1-M_1/M_2)\dot{M_1}/M_1$ under conservative
mass transfer via Roche lobe overflow(Verbunt 1993).  In order for the orbit
not to widen, the mass transfer rate via Roche lobe overflow must be less than
a value of about $10^{-9}-10^{-8}$ $M_\odot/yr$ for a primary of 3-100
$M_\odot$.  In a steady-state, accretion disks with these accretion rates
generate luminosities of $10^{37}-10^{38}$ erg/s regardless of the mass of the
accretor, less than $10^{39}$ erg/s we observed for X-7.  The larger observed
energy would require an additional phenomenon, such as a transient outburst or
relativistic beaming.  For example, X-ray novae can have substantial outbursts,
but usually with a duty cycle of less than 10\%.  However, this source has been
brighter than $10^{39}$ erg/s during the past two decades (Figure 3), in
conflict with expectations for dwarf novae.  Therefore, the more likely
explanation for the high luminosities is by relativistic beaming, as suggested
by King et al. (2001).

An alternative model for the binary and the periodicity is that the period
arises due to a hot spot in the accretion disk around the primary.  This
picture removes the restrictions on the mass of the secondary and its
separation from the primary.  If we assume Keplerian motion of the hot spot
around the accretor, this 2 hour period corresponds to a distance of $\sim1-2$
R$_\odot$ for an accretor mass of $\sim3-10$ M$_\odot$. At such distances, the
multi-color disk temperature is normally $10^4$ K. A hot spot at such distances
fails to explain the fluctuation amplitude for hard photons ($>1.5$ keV).

Another possibility is that the period is not real and results from the limited
sampling of a randomly variable object.  If the period reflects that of orbital
motion, it should be highly reproducible, provided that the source is
sufficiently bright.  If it is due to the motion of a hot spot on a disk, one
might expect the hot spot to fade with time (then the period would vanish while
the source might still be luminous) or the spot might move inward in the disk,
leading to a shorter period.  Additional observations that could distinguish
between these expectations would be greatly beneficial in clarifying the nature
of the variation seen in M51 X-7.

\acknowledgements

We are grateful for the service of Chandra Data Archive. We would like to thank
Renato Dupke, Eric Miller, Rick Rothschild, and Steven Murray for helpful
discussions.  We gratefully acknowledge support for this work from NASA under
grants HST-GO-09073.

\clearpage

\begin{deluxetable}{lllllllll}
\tablecaption{Spectral Fits for M51 X-7 in 2 ACIS observations}
\tabletypesize{\tiny}
\tablehead{
\colhead{ObsID} & \colhead{model} & \colhead{$N_H$} & \colhead{$\Gamma$} &
\colhead{$kT_{in}$} & \colhead{$kT$} & \colhead{$\chi_\nu^2$/dof} & \colhead{absorbed $L_X$\tablenotemark{a}} & \colhead{unabsorbed $L_X$\tablenotemark{a}} \\
\colhead{}& \colhead{}& \colhead{($10^{22}$ $cm^{-2}$)} & \colhead{} &
\colhead{(keV)} & \colhead{(keV)} &\colhead{} & \colhead{($10^{38}$erg/s)} & \colhead{($10^{38}$erg/s)}
}

\startdata
    & wabs*powerlaw & $0.06^{+0.07}_{-0.06}$ & $1.25^{+0.24}_{-0.22}$ & \nodata & \nodata & 0.95/21 & 18.6 & 19.8 \\
354 & wabs*diskbb   & $<0.03$ & \nodata & $2.12^{+0.71}_{-0.42}$ & \nodata & 1.06/21 & 17.1 & 17.1 \\
    & wabs*bb       & \nodata   & \nodata & \nodata & \nodata & 3.47/21 & \nodata & \nodata\\
    & wabs*(bb+diskbb) & $0.02^{+0.09}_{-0.02}$ &\nodata&$0.82^{+1.57}_{-0.30}$ & 2.0 (frozen)& 1.01/20 & 18.8 & 19.2 \\
\hline
    & wabs*powerlaw & $0.70^{+1.02}_{-0.55}$ & $>2.6$ & \nodata & \nodata & 0.82/5  & 0.39 & 9.34\\
1622& wabs*diskbb   & $0.27^{+1.09}_{-0.27}$ & \nodata & $0.29\pm0.09$ & \nodata & 0.89/5  & 0.30 & 0.77 \\
    & wabs*bb       & \nodata                & \nodata & \nodata & \nodata   & 4.61/7  & \nodata & \nodata \\
\enddata
\tablenotetext{a}{The X-ray luminosities are in 0.3-8 keV.}
\end{deluxetable}

\clearpage

\begin{figure}
\plotone{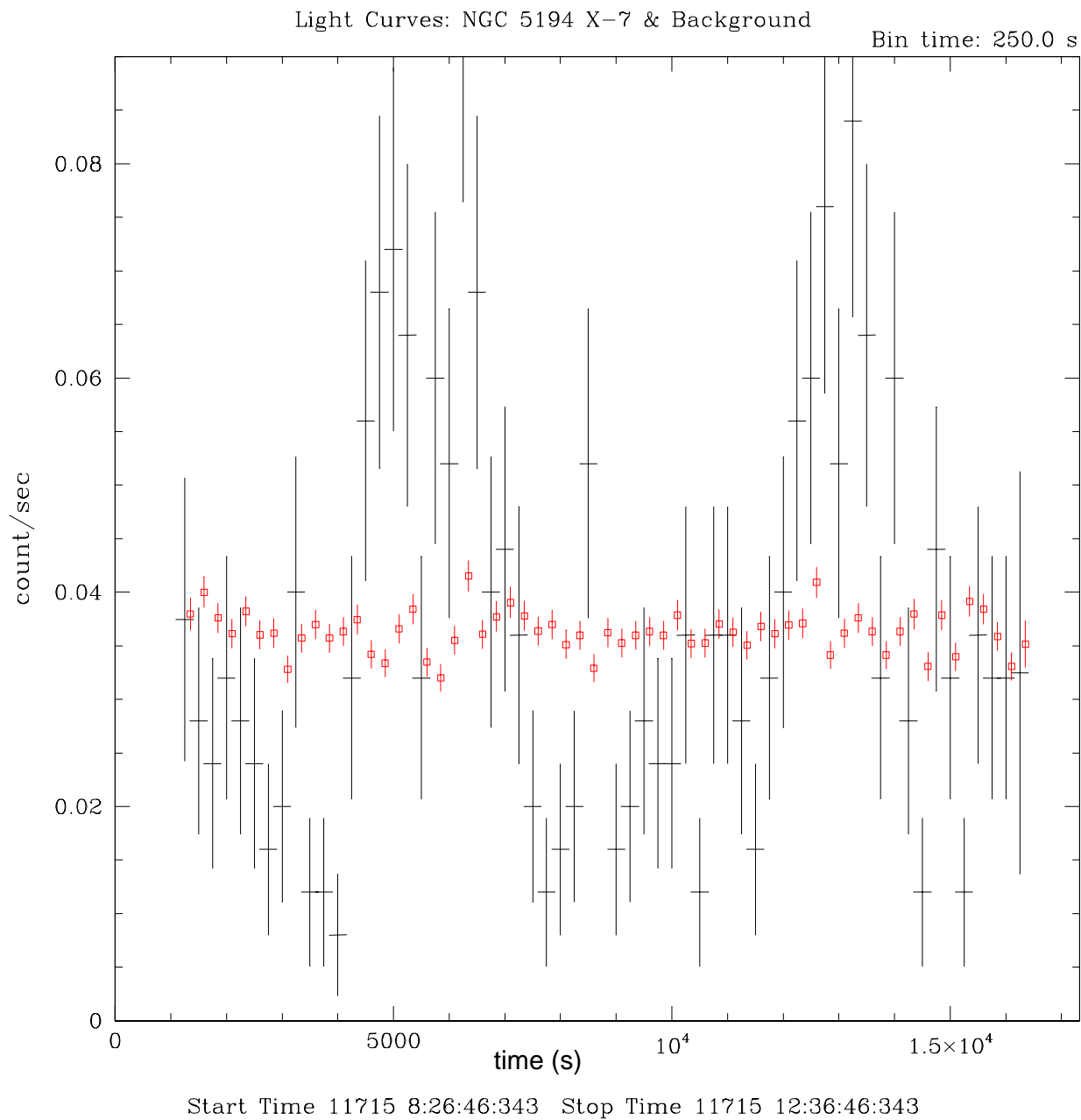}
\caption{Light curve of M51 X-7 in ACIS observation ObsID354. The ULX flux
varies periodically by an amplitude of more than 50\%. The overlaid light curve
labeled by squares is the scaled background, which shows variation of less
than 10\%. Photons within 0.3-10 keV were used for the light curves.}

\end{figure}

\begin{figure}
\epsscale{0.8}
\plotone{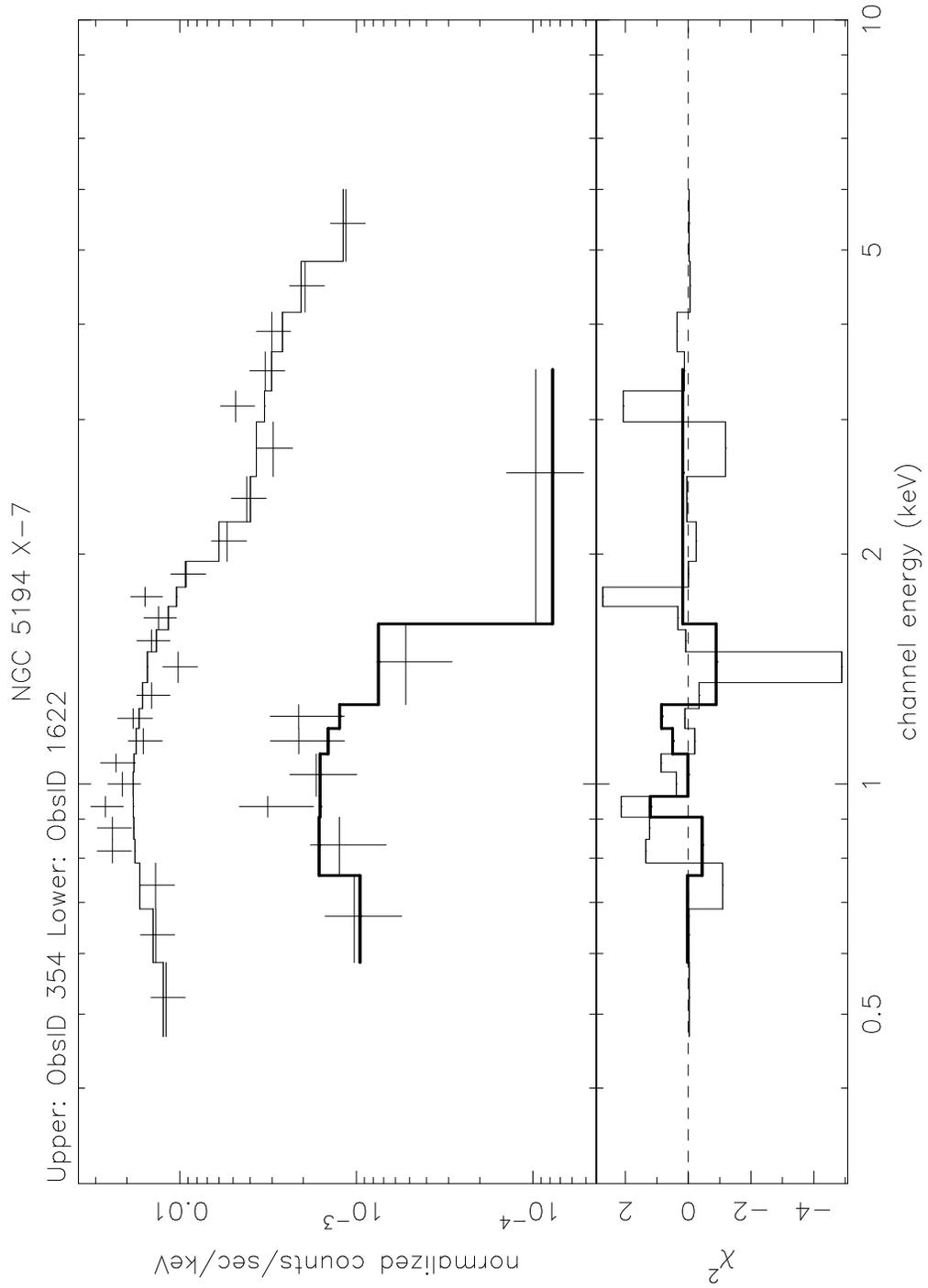}
\caption{Spectral fits of M51 X-7 in two ACIS observations. In the first
observation (ObsID 354) the ULX was in a high hard state, while in the second
observation (ObsID 1622, model and $\chi^2$ in thick lines) it changed to a low soft state.}

\end{figure}

\begin{figure}
\plotone{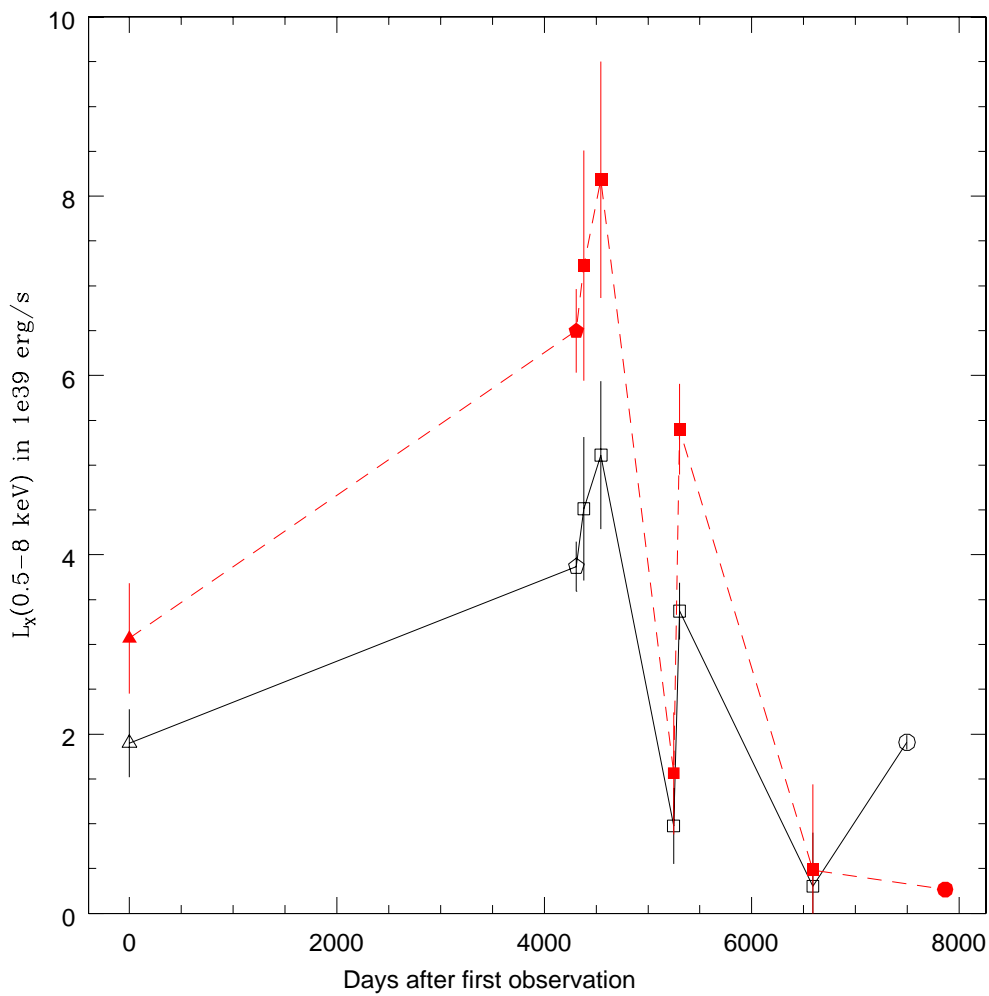}
\caption{Long term variability of M51 X-7. Extracted from the archive were
1 EINSTEIN HRI observation (triangles), 1 ROSAT PSPC observation (pentagons), 5
ROSAT HRI observations (squares) and 2 Chandra ACIS observations (circles). The
instrumental count rates were converted to flux assuming absorbed power-law
models (a) if in the high-hard state (open symbols), and (b) if in the low-soft
state (filled symbols). A distance of 7.7 Mpc was assumed to convert fluxes to
luminosities. }

\end{figure} 

\end{document}